\documentclass[iop]{emulateapj}

\usepackage{atbegshi}

\def\ltsima{$\; \buildrel < \over \sim \;$}
\def\simlt{\lower.5ex\hbox{\ltsima}}
\def\gtsima{$\; \buildrel > \over \sim \;$}
\def\simgt{\lower.5ex\hbox{\gtsima}}
\def\kms{{\rm\,km\,s^{-1}}}

\def\kpc{{\rm\,kpc}}

\def\msun{{\rm\,M_\odot}}
\def\lsun{{\rm\,L_\odot}}

\def\pc{{\rm\,pc}}

\def\AA{$\; \buildrel \circ \over {\rm A}$}

\def\s{\ifmmode \widetilde \else \~\fi}
\def\={\overline}

\def\spose#1{\hbox to 0pt{#1\hss}}

\def\eg{{e.g.,\ }}

\def\lta{\mathrel{\spose{\lower 3pt\hbox{$\mathchar"218$}}
     \raise 2.0pt\hbox{$\mathchar"13C$}}}
\def\gta{\mathrel{\spose{\lower 3pt\hbox{$\mathchar"218$}}
     \raise 2.0pt\hbox{$\mathchar"13E$}}}
\def\Dt{\spose{\raise 1.5ex\hbox{\hskip3pt$\mathchar"201$}}}    
\def\dt{\spose{\raise 1.0ex\hbox{\hskip2pt$\mathchar"201$}}}    

\def\dotsfill{\leaders\hbox to 1em{\hss.\hss}\hfill}

\def\Gyr{{\rm\,Gyr}}
\def\FeH{{\rm[Fe/H]}}

\shorttitle{Spectroscopy of Triangulum~II}
\shortauthors{N. F. Martin et al.}

\begin{document}


\title{Triangulum~II: a very metal-poor and dynamically hot stellar system}


\author{Nicolas F. Martin$^{1,2}$, Rodrigo A. Ibata$^1$, Michelle L. M. Collins$^{3,4,5}$, R. Michael Rich$^6$, Eric F. Bell$^7$, Annette M. N. Ferguson$^8$, Benjamin P. M. Laevens$^{1,2}$, Hans-Walter Rix$^2$, Scott C. Chapman$^9$, Andreas Koch$^{10}$}

\email{nicolas.martin@astro.unistra.fr}

\altaffiltext{1}{Observatoire astronomique de Strasbourg, Universit\'e de Strasbourg, CNRS, UMR 7550, 11 rue de l'Universit\'e, F-67000 Strasbourg, France}
\altaffiltext{2}{Max-Planck-Institut f\"ur Astronomie, K\"onigstuhl 17, D-69117 Heidelberg, Germany}
\altaffiltext{3}{Astronomy Department, Yale University, New Haven, CT 06520}
\altaffiltext{4}{Department of Physics, University of Surrey, Guildford, GU2 7XH, Surrey, UK}
\altaffiltext{5}{Hubble Fellow}
\altaffiltext{6}{Department of Physics and Astronomy, University of California, Los Angeles, PAB, 430 Portola Plaza, Los Angeles, CA 90095-1547, USA}
\altaffiltext{7}{Department of Astronomy, University of Michigan, 500 Church St., Ann Arbor, MI 48109, USA}
\altaffiltext{8}{Institute for Astronomy, University of Edinburgh, Royal Observatory, Blackford Hill, Edinburgh EH9 3HJ, UK}
\altaffiltext{9}{Department of Physics and Atmospheric Science, Dalhousie University, Coburg Road, Halifax, NS B3H 1A6, Canada}
\altaffiltext{10}{Landessternwarte, Zentrum f\"ur Astronomie der Universit\"at Heidelberg, K\"onigstuhl 12, 69117 Heidelberg, Germany}

\begin{abstract}
We present a study of the recently discovered compact stellar system Triangulum~II. From observations conducted with the DEIMOS spectrograph on Keck~II, we obtained spectra for 13 member stars that follow the CMD features of this very faint stellar system and include two bright red giant branch stars. Tri~II has a very negative radial velocity ($\langle v_r\rangle=-383.7^{+3.0}_{-3.3}\kms$) that translates to $\langle v_{r,\mathrm{gsr}}\rangle\simeq-264\kms$ and confirms it is a Milky Way satellite. We show that, despite the small data set, there is evidence that Tri~II has complex internal kinematics. Its radial velocity dispersion increases from $4.4^{+2.8}_{-2.0}\kms$ in the central $2'$ to $14.1^{+5.8}_{-4.2}\kms$ outwards. The velocity dispersion of the full sample is inferred to be $\sigma_{vr}=9.9^{+3.2}_{-2.2}\kms$. From the two bright RGB member stars we measure an average metallicity $\langle\FeH\rangle=-2.6\pm0.2$, placing Tri~II among the most metal-poor Milky Way dwarf galaxies. In addition, the spectra of the fainter member stars exhibit differences in their line-widths that could be the indication of a metallicity dispersion in the system. All these properties paint a complex picture for Tri~II, whose nature and current state are largely speculative. The inferred metallicity properties of the system however lead us to favor a scenario in which Tri~II is a dwarf galaxy that is either disrupting or embedded in a stellar stream.
\end{abstract}


\section{Introduction}
A large number of faint and small stellar systems have been uncovered over the last decade thanks to wide photometric surveys. The harvest of such objects, which started with Willman~1 \citep{willman05a} and then Segue~1 \citep{belokurov07a}, blossomed through systematic searches of the Sloan Digital Sky Survey \citep[SDSS;][]{belokurov09} and, more recently, of the Dark Energy Survey \citep[DES;][]{bechtol15,drlica-wagner15,koposov15} and the Panoramic Telescope and Rapid Response System~1 \citep[Pan-STARRS1;][]{laevens15a,laevens15b}. The photometric properties of many of these systems ambiguously locate them in a region of parameter space where dwarf galaxies appear to mix with globular clusters \citep{gilmore07}. Spectroscopic studies of their stars are therefore unavoidable to show that they are either dynamically cold and display no metallicity dispersion \citep[\eg Laevens~1;][]{kirby15a}, as expected for globular clusters, or that they are dynamically hot \citep[\eg Segue~1;][]{geha09}, have a metallicity dispersion \citep[\eg Segue~2;][]{kirby13b}, and/or lie on the luminosity--metallicity relation followed by dwarf galaxies \citep[\eg Hydra~II and Draco~II;][]{kirby15a, martin15b}.

Although they are expected by the dozen in simulations \citep{tollerud08,bullock10}, only a handful of these faint and small stellar systems have so far been confirmed as dwarf galaxies. Any new addition to the list is particularly valuable as these objects are among the most promising for the indirect detection of the elusive dark matter particle \citep[\eg][]{geringer-sameth15}. Their small baryonic component ($L\sim10^{2-4}\lsun$) makes them hard to find and study but, at the same time, gives powerful insight into the interplay of physical processes that drive galaxy formation at low masses and in shallow potential wells. The characterization of these systems is, however, made difficult by the potential presence of binary stars that can significantly inflate the intrinsic velocity dispersion of a system with a velocity dispersion of only a few $\kms$ \citep{mcconnachie10}. In addition, the usual assumption of dynamical equilibrium can be inappropriate for systems that are often found within $\sim40\kpc$ of the Galactic center, further impeding their study \citep[\eg Willman~1;][]{willman11}. Finally, the difficulty to disentangle member stars from foreground contamination can sometimes further compound the analysis of these faint objects \citep[\eg][]{bonnivard15}.

In this paper, we report a spectroscopic study of the Triangulum~II (Tri~II) stellar system discovered by \citet{laevens15a} in Pan-STARRS1 and confirmed with deep Large Binocular Camera (LBC) photometry. Tri~II is very faint ($M_V=-1.8\pm0.5$), fairly compact ($r_h=34^{+9}_{-8}\pc$), and located at $30\pm2\kpc$ from the Sun, or $36\pm2\kpc$ from the Galactic center. So far as one can infer from the photometry, it appears to contain only old and metal-poor stars.

We present the Keck~II/DEIMOS data used for the analysis in section~2 of this paper, the results of the spectroscopic study in section~3, while section~4 is devoted to a discussion on our findings.

\section{Observations and data}
Two masks targeting Tri~II potential member stars were observed during the night of September 17, 2015, with the DEep Imaging Multi-Object Spectrograph (DEIMOS) on Keck~II \citep{faber03} under reasonably good conditions (0.7--1.0$"$ seeing and $\sim60$\% humidity; PI: Rich, program ID: 2015B\_U064D). The LBC photometry used by \citet{laevens15a} to confirm the discovery of the satellite were used to place slits on stars selected in the color-magnitude diagram (CMD) to follow the system's main sequence turn off (MSTO), sub-giant branch (SGB), and red giant branch (RGB). The selection was purposefully tight around the sharp MSTO and loose around the RGB region that cannot be selected out of the MW foreground contamination from the photometry alone.

Each mask was integrated for 3600s, split into 3 sub-exposures for cosmic-ray removal. The spectrograph was set up with the 1200 lines/mm grating, which translates to $\sim0.33$\AA\ per pixel in the Calcium \textsc{ii} triplet (CaT) region we focus on. The full spectra cover the range 6600--9400\AA. Raw frames are processed through our own pipeline, which we developed over the years to specifically reduce DEIMOS spectra. We refer the reader to \citet{ibata11a} for an overview of the details of the processing, and the results of using the pipeline on high quality data. Briefly, the reduction method calibrates each pixel of the original spectroscopic frame in both wavelength and spatial position on the sky. In this way the data retain their original pixel binning, and one avoids introducing the correlated noise that occurs when spectra are extracted and co-added. The wavelength solution is given by a fit to arc-lamp frames taken immediately after the science frames. We also allow for a wavelength re-calibration using the Fraunhofer A band in the range 7595--7630\AA\ in order to perform small telluric corrections when the signal to noise of the spectra is sufficient. A two-dimensional sky spectrum model is built for each slitlet following a procedure inspired by the method of \citet{kelson03}. Finally, the radial velocity (and corresponding uncertainty) of the target stars is measured by fitting a simple Gaussian model of the Ca~II triplet lines to the pixel data minus the sky spectrum. Only stars with large enough signal-to-noise ($S/N>3$ per pixel) and velocity uncertainties lower than $15\kms$ are kept for the analysis, leaving a total sample of 50 stars.

Velocity uncertainties measured by the pipeline are known not to fully account for low level systematics. Following \citet{ibata11a}, we add an uncertainty floor of $2.25\kms$ in quadrature to the velocity uncertainties measured directly from the spectra. Finally, we measure the equivalent widths of the CaT lines and their uncertainties by independently fitting to the three lines Moffat functions shifted to the velocity of a given star \citep{ibata05}.

Unless specified otherwise, radial velocities reported in this paper are heliocentric velocities, corrected from the motion of the Earth around the Sun, but not corrected for the Solar motion. The known and derived properties of Tri~II are summarized in Table~\ref{summary} and the properties of the spectroscopic sample stars are listed below in Table~\ref{data}.

\begin{table}
\caption{\label{summary}Properties of Tri~II}
\begin{tabular}{l|c}
RA\footnote{from \citet{laevens15a}} (ICRS) & 02:13:17.4\\
Dec$^\mathrm{a}$ (ICRS) & +36:10:42.4\\
Heliocentric distance$^\mathrm{a}$ (\kpc) & $30\pm2$\\
Galactocentric distance$^\mathrm{a}$ (\kpc) & $36\pm2$\\
$r_h$$^\mathrm{a}$ ($'$) & $3.9^{+1.1}_{-0.9}$\\
$r_h$$^\mathrm{a}$ ($\pc$) & $34^{+9}_{-8}$\\
$M_V$$^\mathrm{a}$ & $-1.8\pm0.5$\\
$L_V (\lsun)$$^\mathrm{a}$ & $10^{2.6\pm0.2}$\\
\hline
 & Global kinematics\\
$\langle v_r \rangle$ ($\kms$) & $-383.7^{+3.0}_{-3.3}$\\
$\langle v_{r,gsr} \rangle$ ($\kms$) & $-264$\\
$\sigma_{vr}$ ($\kms$) & $9.9^{+3.2}_{-2.2}$\\
\hline
 & Inner kinematics ($<2'$)\\
$\langle v_r \rangle$ ($\kms$) & $-379.8^{+2.1}_{-2.7}$\\
$\sigma_{vr}$ ($\kms$) & $4.4^{+2.8}_{-2.0}$\\
\hline
 & Outer kinematics ($>2'$)\\
$\langle v_r \rangle$ ($\kms$) & $-387.3^{+5.7}_{-6.3}$\\
$\sigma_{vr}$ ($\kms$) & $14.1^{+5.8}_{-4.2}$\\
\hline
$\langle\FeH\rangle$ & $-2.6\pm0.2$\\
\end{tabular}
\end{table}
 
\section{Analysis}
\subsection{Velocities}
\begin{figure}
\begin{center}
\includegraphics[width=0.75\hsize,angle=270]{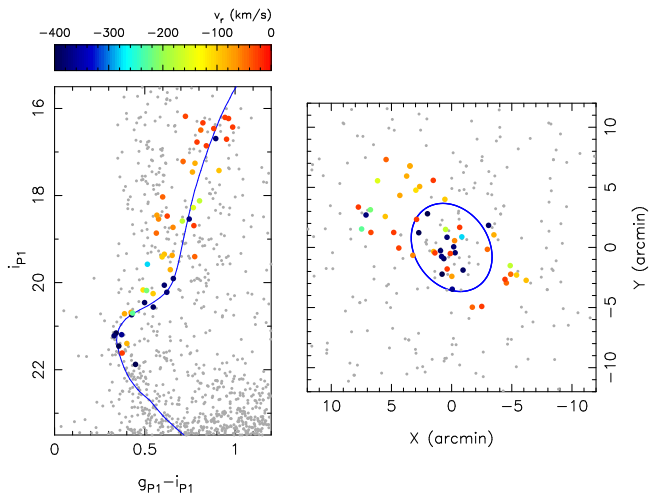}
\caption{\label{CMD}\emph{Left:} LBC CMD of stars within $2r_h$ of Tri~II's centroid. Stars with spectra that passed our quality cuts are shown color-coded by velocity whereas stars without spectroscopic information are represented by gray dots. The 13 Tri~II member stars appear as dark blue points with large negative velocities and follow the \textsc{Parsec} isochrone \citep{bressan12} shown in blue, favored by \citet{laevens15a} to reproduce the CMD features of the stellar system ($13\Gyr$ and $\FeH=-2.2$). \emph{Right:} Distribution of the LBC stars selected to follow the Tri~II CMD features. The color-coding is the same as in the left-hand panel. The blue ellipse represents the region within the half-light radius of Tri~II.}
\end{center}
\end{figure}

Stars with good quality spectra are displayed in Figure~\ref{CMD} over the CMD of Tri~II in the left-hand panel and over the spatial distribution of possible Tri~II stars in the right-hand panel. A group of stars with highly negative velocities, shown in dark-blue, is almost perfectly aligned with the favored old and metal-poor isochrone of \citet[][$13\Gyr$ and $\FeH=-2.2$]{laevens15a}. Most of these stars are MSTO or SGB stars but the sample also contains 2 RGB stars that shall prove valuable to derive the metallicity of Tri~II. A large fraction of the stars with very negative velocities is also concentrated within the half-light radius of Tri~II represented by the blue ellipse, even though some member stars are also located beyond and throughout the region covered by the DEIMOS masks.

\begin{figure}
\begin{center}
\includegraphics[width=0.75\hsize,angle=270]{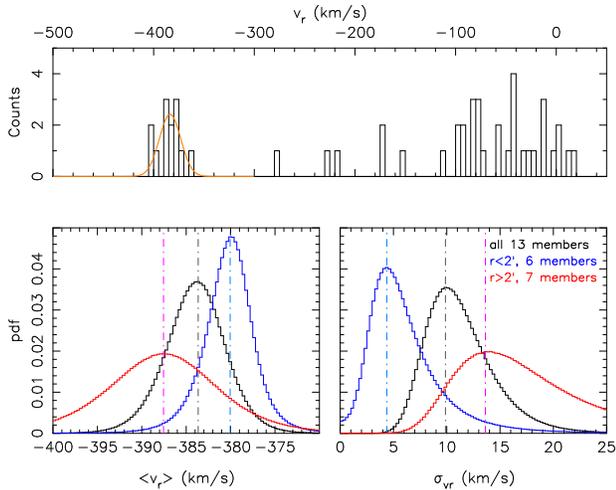}
\caption{\label{velFit}\emph{Top:} Heliocentric velocity distribution of the spectroscopic sample. The velocity peak of Tri~II stars is visible at $v_r\sim-385\kms$, separated from the MW contamination with $v_r>-300\kms$. The orange line displays the best fit to the velocity distribution of Tri~II stars, convolved by the median velocity uncertainty. \emph{Bottom:} Pdfs of the systemic velocity (left) and velocity dispersion (right) of the full Tri~II sample (black histograms). The blue and red histograms correspond to the pdfs for the inner and outer half of the sample, respectively. The vertical lines indicate the modes of the distributions. Note the discrepant velocity dispersion pdfs for the inner and outer samples.}
\end{center}
\end{figure}

The velocity distribution of the sample stars shown in the top panel of Figure~\ref{velFit} reveals the velocity peak produced by Tri~II stars. Located around $v_r\sim-385\kms$, it appears completely isolated from the MW foreground contamination and confirms that all the stars shown in dark blue in Figure~\ref{CMD} are member stars. The velocity peak is also surprisingly wide. Fitting a Gaussian distribution to the 13 member stars corroborates this first impression as we infer a velocity dispersion of $\sigma_{vr}=9.9^{+3.2}_{-2.2}\kms$ around a systemic velocity of $\langle v_r\rangle=-383.7^{+3.0}_{-3.3}\kms$ (see the bottom panels of Figure~\ref{velFit} for the parameters' probability distribution functions or pdfs). Such a value is at odds with velocity dispersion measurements usually obtained in similarly faint and compact MW systems. Recent studies consistently infer values of, at most, a few $\kms$ \citep{martin07a,simon07,geha09,willman11,kirby13,kirby15a,martin15b}.

\begin{figure}
\begin{center}
\includegraphics[width=0.7\hsize,angle=270]{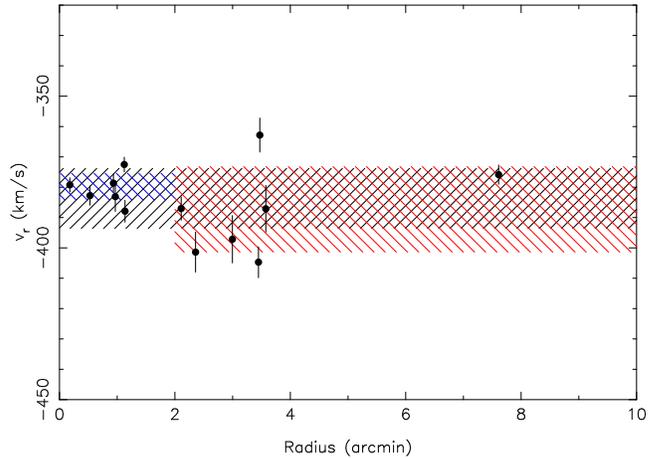}
\caption{\label{R_vel}Velocities of Tri~II member stars as a function of their distance from the system's centroid, showing an apparent flaring of the radial velocity distribution with distance. The hashed regions correspond to the velocities within $\pm\sigma_{vr}$ for the full sample (gray), the inner sample (blue), and the outer sample (red).}
\end{center}
\end{figure}

Figure~\ref{R_vel} shows, however, that the radial velocities of member stars appear to flare up with distance from the center of the system. The central half of the sample is much more closely aligned in velocity than its outer half. Fitting the velocity distribution for the 6~member stars within $2'$ yields $\langle v_r\rangle=-379.8^{+2.1}_{-2.7}\kms$ and $\sigma_{vr}=4.4^{+2.8}_{-2.0}\kms$, whereas the outermost 7 stars yield $\langle v_r\rangle=-387.3^{+5.7}_{-6.3}\kms$ and $\sigma_{vr}=14.1^{+5.8}_{-4.2}\kms$. While the systemic velocities of the inner and outer samples are compatible, this is hardly the case for the velocity dispersion measurements (see pdfs in the bottom panels of Figure~\ref{velFit}). Although we cannot completely rule out the compatibility of the two measurements inferred from a small number of stars in both samples, we find that the velocity dispersions are nevertheless discrepant at the $2\sigma$ level, confirming the visual impression from Figure~\ref{R_vel}.

The large velocity dispersion of the outer sample is robust to the velocity uncertainties as it remains present, even if we increase the uncertainty floor to the unlikely value of $4\kms$. We also checked for the presence of a velocity gradient as a function of position that could artificially give the impression of the flaring from stars with significantly different velocities on opposite sides of the system. No such gradient was found in the data.

\subsection{Metallicities}
\begin{figure}
\begin{center}
\includegraphics[width=\hsize]{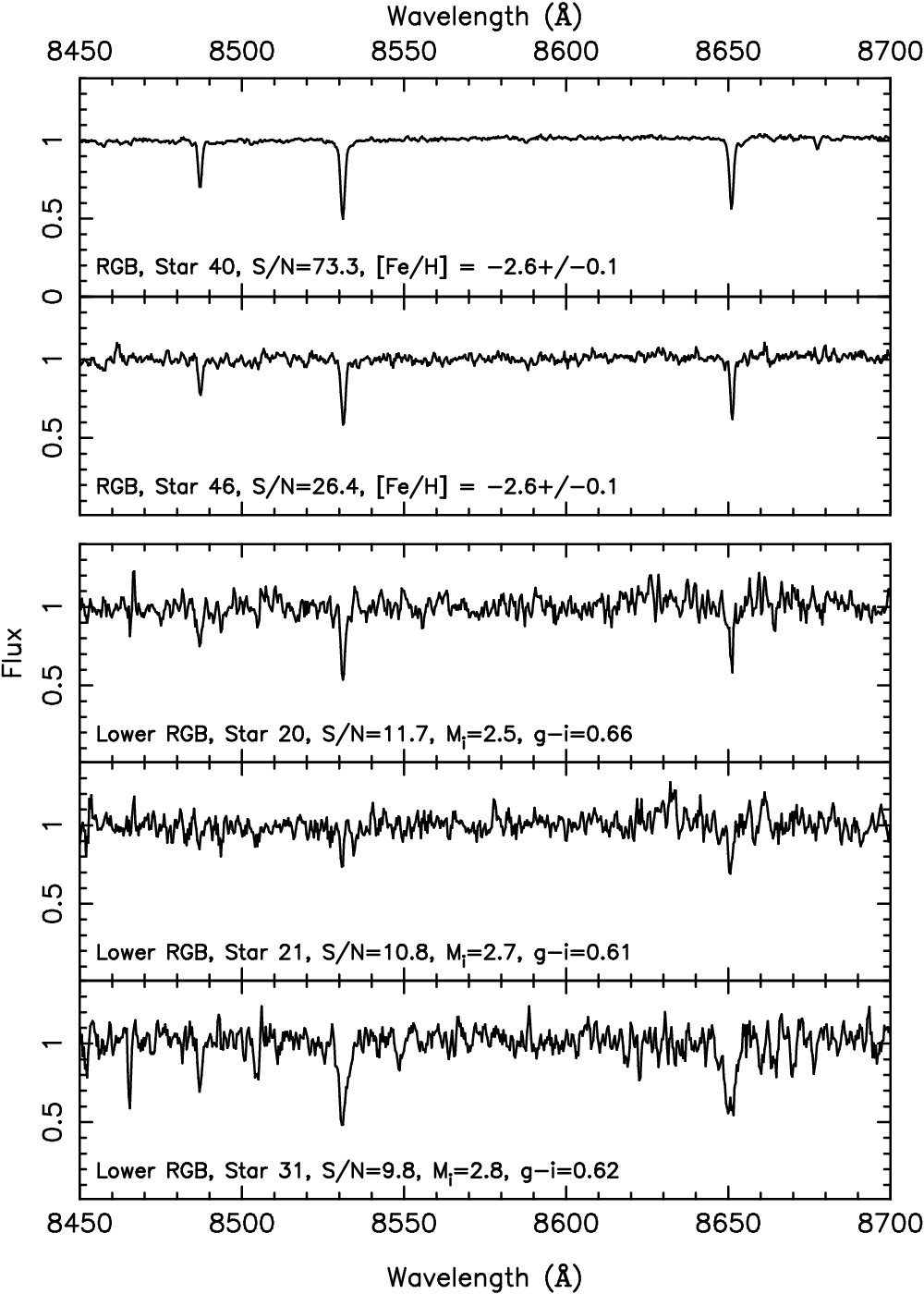}
\caption{\label{spectra} Smoothed spectra of two Tri~II RGB (top) and three lower-RGB/SGB member stars in the region around the CaT. The spectra are smoothed with a 3-pixel boxcar filter. The properties of the stars are listed in each panel. Note the weak lines of the two RGB stars and the varying lines widths of the three lower-RGB/SGB stars that share similar photometric properties, indicating a potential metallicity dispersion in the system.}
\end{center}
\end{figure}

Two of the observed member stars (stars 40 \& 46 in Table~\ref{data}) are RGB stars bright enough to allow for a determination of their $\FeH$ metallicity via a measure of the equivalent widths of the CaT lines. \citet{starkenburg10} has calibrated the relation between the equivalent widths of the second and third Ca\textsc{ii} lines, $EW_{2+3}$, down to very low metallicities ($-4.0<\FeH<-0.5$). We first convert the $i_\mathrm{P1}$ magnitudes of the two stars to $I_c$ magnitudes with the \citet{tonry12} color equation and, using the \citet{starkenburg10} relation, we calculate $\FeH=-2.6\pm0.1$ for both RGB stars. These very low metallicity values are confirmed by the inspection of the spectra (top two panels of Figure~\ref{spectra}) that both exhibit very weak CaT lines. Assuming a Gaussian metallicity distribution function, we infer a mean metallicity $\langle\FeH\rangle=-2.6\pm0.2$ for Tri~II, which is therefore among the most metal-poor MW satellites.

It is harder to provide a definite conclusion on the presence or absence of a metallicity dispersion in Tri~II since two stars alone cannot rule out the presence of a dispersion, even if they are measured to have the same metallicity. Moreover, the other observed member stars are located far below the horizontal branch of the system and the horizontal branch marks the faint limit to which the \citet{starkenburg10} relation has been calibrated. Finally, directly extracting a measure of Fe line strengths from DEIMOS spectra with $S/N\lta10$ is fraught with peril.

We note however that the three Tri~II stars above the SGB that have similar colors ($g_\mathrm{P1}-i_\mathrm{P1}\sim0.65$), magnitudes ($i_\mathrm{P1}\sim20.1$), and signal-to-noise ($S/N\sim10$), have inconsistent equivalent widths with $EW_{2+3}=1.41 \pm 0.21$\AA, $0.87 \pm 0.20$\AA, and $3.90 \pm 0.45$\AA\ from the brighter to the fainter star (stars \# 20, 21, and 31 from Table~\ref{data}, respectively). Moreover, none of these stars show strong Na\textsc{i} doublet lines (8183 and 8192\AA) that would indicate that they are foreground contaminants. These $EW_{2+3}$ differences are directly visible on the spectra (lower three panels of Figure~\ref{spectra}) and could be interpreted as evidence of a metallicity dispersion in Tri~II. Blindly applying the Starkenburg et al. (2010) relation
for these stars that are fainter than the magnitude range over which it was calibrated\footnote{However, see \citet{leaman13} and their study of the metal-poor globular cluster M15 for which the calibration is shown to hold $\sim2$ magnitudes below this system's horizontal branch.} yields tentative metallicities of $\FeH=-2.3\pm0.2$, $-3.0\pm0.6$, and $-0.7\pm0.2$, respectively. From the 5 brightest confirmed Tri~II members, we infer $\langle\FeH\rangle = -2.2^{+0.4}_{-0.3}$ and a large metallicity dispersion of $\sim0.8$ dex. We nevertheless stress that direct $\FeH$ measurements are needed from higher $S/N$ spectra to bolster this marginal conclusion.

\section{Discussion}
We obtained spectra for 13 member stars in the very faint MW satellite Tri~II. These stars follow the CMD features of the stellar system and include mainly MSTO and SGB stars, as well as two bright RGB stars. With $\langle v_r\rangle=-383.7^{+3.0}_{-3.3}\kms$, Tri~II has a very negative radial velocity that translates to $\langle v_{r,\mathrm{gsr}}\rangle\simeq-264\kms$. We have further shown that, as far as we can tell from only 13 member stars, the internal kinematics of Tri~II appear complex with evidence for a radial velocity dispersion increase from $4.4^{+2.8}_{-2.0}\kms$ in the central $2'$ to $14.1^{+5.8}_{-4.2}\kms$ outwards. The velocity dispersion of the full sample is inferred to be $\sigma_{vr}=9.9^{+3.2}_{-2.2}\kms$. Finally, the two bright RGB member stars are both measured to have $\FeH=-2.6\pm0.1$ and point to Tri~II being among the most metal-poor MW satellites. The spectra of fainter member stars exhibit differences in their line-widths that could be due to a metallicity dispersion in the system.

At the distance of Tri~II ($36\pm2\kpc$ from the Galactic center), such a fast infalling velocity is not unexpected for a satellite bound to the MW. It does, however, rule out any association with the numerous stellar structures found nearby in the Milky Way halo. TriAnd \citep{rocha-pinto04}, TriAnd~2 \citep{martin07b}, or the PAndAS MW stream \citep{martin14a} all have positive $v_{r,\mathrm{gsr}}$ \citep{deason14}. Tri~II is also unrelated to the Segue~2 satellite that is located only $\sim10\kpc$ away but has a very different velocity \citep[$\langle v_r\rangle=-39.2\pm2.5\kms$;][]{belokurov09}.

But what is the nature of Tri~II? Taken at face value, the large global velocity dispersion, the very low-metallicity, and the potential metallicity dispersion seem to point towards Tri~II being a dwarf galaxy rather than a globular cluster. However, the complex kinematics of the system question the assumption of dynamical equilibrium that is required to translate a large velocity dispersion into a large mass and mass-to-light ratio. 

\begin{figure}
\begin{center}
\includegraphics[angle=270,width=\hsize]{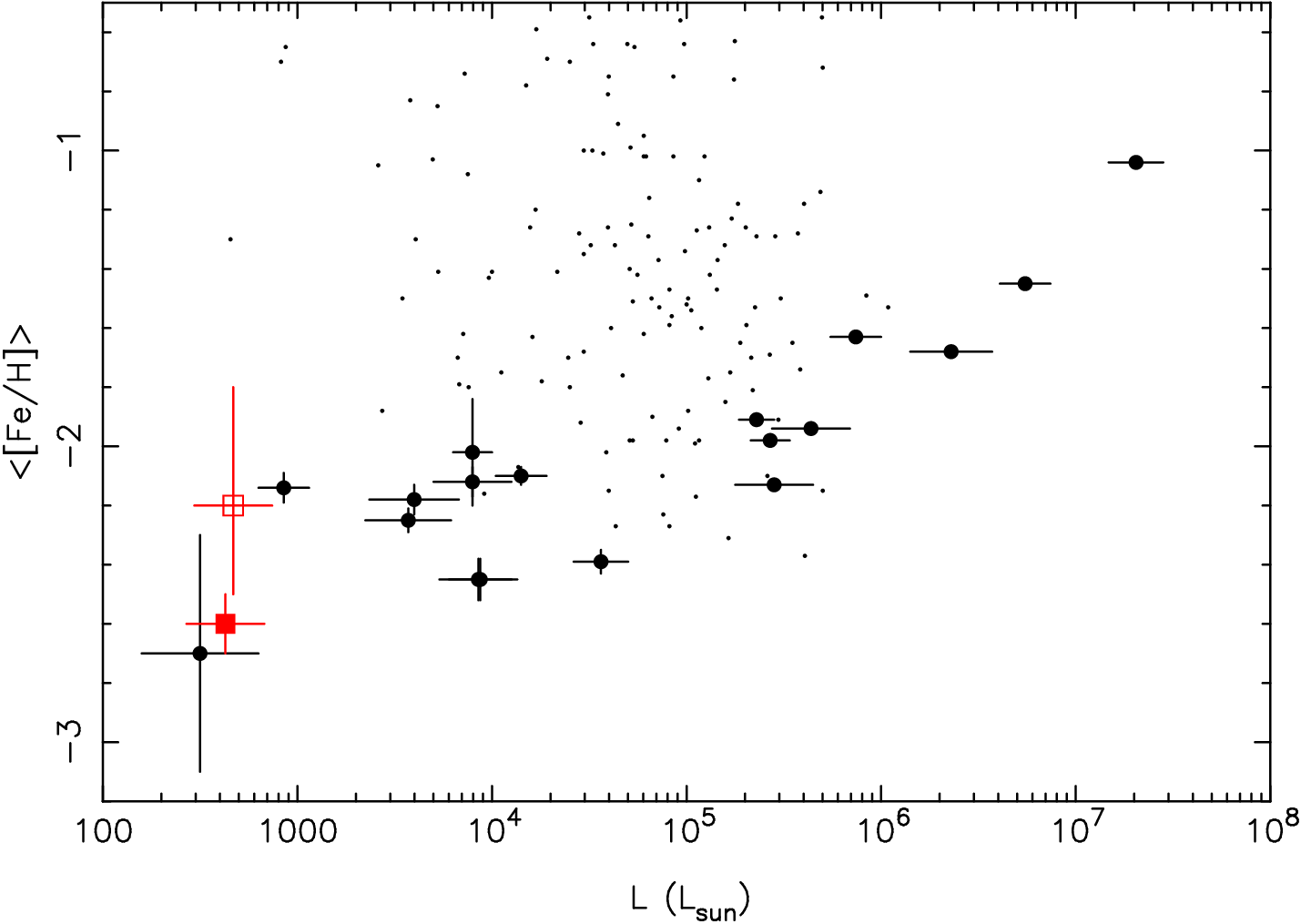}
\caption{\label{FeH_L}Distribution of MW satellites in the mean metallicity versus luminosity plane. Large black points correspond to MW dwarf galaxies, as listed in \citet{kirby13}, supplemented by \citet{norris10}, \citet{kirby15a}, and \citet{simon15}. MW globular clusters are shown as small black dots \citep{harris96}. The red squares corresponds to the Tri~II measurements, with the filled square representing the inference from the two robust individual stellar metallicities (stars 40 and 46) while the hollow circle corresponds to the inference from the 5 stars shown in Figure~\ref{spectra} (the two Tri~II points have been slightly offset from each other along the luminosity axis so their error bars do not overlap). Both measurements compare well with those of other MW dwarf galaxies.}
\end{center}
\end{figure}

\emph{Can $\FeH$ discriminate between globular cluster and dwarf galaxy?} Irrespective of its velocity dispersion, Tri~II is among the most metal-poor systems known. No globular cluster is known with a metallicity below $\FeH=-2.4$ \citep{harris96} and only the Segue~1, Bootes~II, and Reticulum~II dwarf galaxies, with whom Tri~II shares many similar properties (total luminosity, size, distance), are as metal-poor with $\FeH=-2.7\pm0.4$ \citep{norris10} or $\FeH\sim -2.5$ \citep{simon11} for Segue~1, $\FeH=-2.9\pm0.2$ \citep{koch14} for Bootes~II, and $\FeH=-2.65\pm0.07$ \citep{simon15} or $\FeH=-2.6\pm0.3$ \citep{walker15} for Reticulum~II. Figure~\ref{FeH_L} shows that Tri~II is in agreement with the dwarf galaxy metallicity--luminosity relation of \citet{kirby13}, even if we include the 3 stars with tentative metallicity measurements for the inference of the mean metallicity (hollow red square). By analogy, the metallicity of the system therefore appears to favor the dwarf galaxy hypothesis, which would be bolstered further by the marginal evidence of a metallicity dispersion.

\begin{figure}
\begin{center}
\includegraphics[angle=270,width=\hsize]{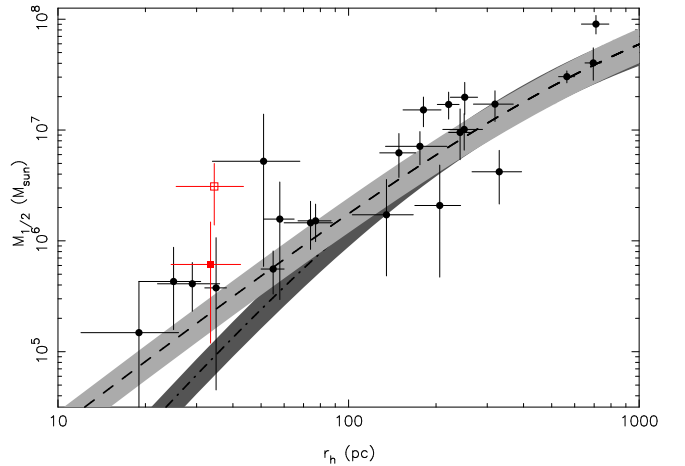}
\caption{\label{mass}Distribution of MW dwarf galaxies in the $r_h$--$\sigma_{vr}$ plane, compared to the Local Group dwarf galaxy mass profiles. The black dashed and dot-dashed lines correspond to the favored Local Group mass profiles of \citet{collins14} for a cored or NFW model, respectively, and the gray bands represent the model dispersions determined by these authors. Black points correspond to Milky Way dwarf galaxies, as listed in \citet{mcconnachie12}, \citet{kirby15a}, \citet{martin15b}, and \citet{simon15}. The Tri~II data point is shown as a hollow square for the global kinematics determined in this paper, or as a filled square when using the kinematics of the inner sample. The former is a strong outlier whereas the latter fits well with other MW dwarf galaxies. The two Tri~II points have been slightly offset from each other along the $r_h$ axis so their error bars do not overlap.}
\end{center}
\end{figure}

\emph{What is the dynamical mass of Tri~II?} It is hard to tell if one takes the increase in the velocity dispersion with radius as a sign that the system is out of equilibrium. On the other hand, if one assumes that the change in $\sigma_{vr}$ is an (unlikely $2\sigma$) statistical fluctuation and that the global velocity dispersion measured at $\sigma_{vr}=9.9^{+3.2}_{-2.2}\kms$ is representative of the true properties of Tri~II, one can easily note that it is a significant outlier among other dwarf galaxies of this size, as displayed in Figure~\ref{mass}. Equation~(1) of \citet{wolf10}, relates the mass within the 3-dimensional half-light radius, $M_{1/2}$, to the half-light radius and velocity dispersion of a bound system in equilibrium and, in the case of Tri~II, yields $M_{1/2}\sim3\times10^6\msun$ and $(M/L)_{1/2}\sim15\,500$ in Solar units. It would mean that Tri~II is almost an order of magnitude more massive than Segue~1 or Reticulum~II and, by a wide margin, the most dark matter system known in the universe. Such a large mass seems very improbable.

\emph{Could MW contaminants pollute the velocity peak?} This also appears unlikely as the inflated velocity dispersion in the outskirts of Tri~II is not driven by any single outlier (see Figure~\ref{R_vel}). The very negative systemic velocity of Tri~II also means it is improbable for the spectroscopic data set to contain more than a single contaminating halo star, if any.

\emph{Is Tri~II disrupting and/or embedded in a stellar stream?} After ruling out that Tri~II is in equilibrium or contaminated by MW halo stars, the most likely hypothesis is that the observed increase in the velocity dispersion with radius is genuine. The large velocity dispersion beyond $2'$ would then be produced by stars that are not bound to the body of the satellite. This seems at odds with the current measure of the half-light radius of Tri~II ($r_h=3.9^{+1.1}_{-0.9}$$'$; \citealt{laevens15a}), but this measure could be systematically biased by the comparatively small LBC field of view \citep{munoz12}. In fact, the spatial distribution of possible satellite member stars exhibits what could be a more compact central core within $\sim2'$ surrounded by a more diffuse component (see the right-hand panel of Figure~1 from \citealt{laevens15a}). Upcoming wider and deeper photometric data will allow us to robustly investigate this morphology.

If this $\sim2'$ core really is the true extent of the main body of Tri~II, the velocity dispersion we measure from the inner half of the spectroscopic sample ($\sigma_{vr}=4.4^{+2.8}_{-2.0}\kms$) would be more representative of the satellite's intrinsic properties and would yield a consistent picture with the very faint dwarf galaxies Segue~1 \citep{simon11}, Reticulum~II \citep{simon15,walker15}, or Draco~II \citep{martin15b}, as can be seen in Figure~\ref{mass}. The only conundrum would then be whether Tri~II is disrupted by tidal interactions with the Milky Way, or if it is in equilibrium but embedded in a stellar stream. Such a stellar stream could, for instance, be produced by a potentially more massive dwarf galaxy it would have been a satellite of in the past \citep[\eg][]{wheeler15}. The large negative velocity of the system however likely rules out a system that has just now been tidally disrupted after a pericentric passage. The fact that the star with the most discrepant spectrum among those shown in Figure~\ref{spectra} (star 31) is also the only one beyond $2'$ could point towards the latter hypothesis but this is hardly conclusive.

At the moment, the puzzling properties of Tri~II mean that its nature and current state are largely speculative. We favor the scenario in which Tri~II is a dwarf galaxy that is either disrupting or embedded in a stellar stream but cannot completely rule out that it could be a disrupting globular cluster. Whatever the true nature of the satellite, it exhibits unexpected properties that make it very exciting and call for more observations to understand its complexity.

%
%
%
%

\begin{table*}
\caption{\label{data}Properties of observed stars meeting the quality criteria}
\begin{tabular}{lllccccccccc}
\# & RA & Dec & R\footnote{Distance from Tri~II's centroid} ($'$) & $g_\mathrm{P1}$ & $i_\mathrm{P1}$ & $v_r$ & $\delta_{vr}$ & S/N & Member? & EW$_{2+3}$ & $\FeH$\\
 & (ICRS) & (ICRS) & ($'$) &  &  & ($\kms$) & ($\kms$) & (per pixel) &  & (\AA) & \\
\hline
 1 &    33.1946678 &    36.1328316 &  6.8 & 19.955 & 19.347 & $ -90.6$ & 2.8 & 23.2 & N & \\
 2 &    33.2111244 &    36.1401672 &  5.9 & 20.006 & 19.410 & $ -99.3$ & 2.6 & 13.2 & N & \\
 3 &    33.4113731 &    36.2505836 &  6.1 & 20.026 & 19.375 & $ -76.8$ & 2.6 & 15.5 & N & \\
 4 &    33.3050842 &    36.1931953 &  1.2 & 20.092 & 19.578 & $-276.2$ & 2.9 & 13.4 & N & \\
 5 &    33.3774185 &    36.2628899 &  5.7 & 20.344 & 19.702 & $ -83.2$ & 2.8 & 12.0 & N & \\
 6 &    33.3839149 &    36.2578049 &  5.6 & 20.658 & 20.168 & $-111.7$ & 3.8 &  7.3 & N & \\
 7 &    33.3341675 &    36.2451935 &  4.0 & 20.799 & 20.253 & $ -97.3$ & 3.5 &  7.1 & N & \\
 8 &    33.2591667 &    36.2090836 &  3.6 & 21.165 & 20.737 & $-387.1$ & 7.7 &  4.9 & Y & \\
 9 &    33.3639183 &    36.2251930 &  3.4 & 21.499 & 21.158 & $-404.7$ & 5.1 &  3.8 & Y & \\
10 &    33.4349174 &    36.3003044 &  9.1 & 20.180 & 19.403 & $ -42.2$ & 4.1 &  7.0 & N & \\
11 &    33.3940010 &    36.2913055 &  7.6 & 21.106 & 20.682 & $ -78.6$ & 4.6 &  4.8 & N & \\
11 &    33.3940010 &    36.2913055 &  7.6 & 21.101 & 20.714 & $ -78.6$ & 4.6 &  4.8 & N & \\
12 &    33.2499161 &    36.1958885 &  3.7 & 18.036 & 17.258 & $ -80.8$ & 2.3 & 48.1 & N & \\
13 &    33.2220421 &    36.1532211 &  5.1 & 19.294 & 18.586 & $-172.5$ & 2.4 & 26.5 & N & \\
14 &    33.2612076 &    36.1757774 &  3.0 & 19.427 & 18.863 & $ -46.2$ & 2.5 & 19.4 & N & \\
15 &    33.3087502 &    36.2062492 &  1.8 & 17.416 & 16.429 & $   1.3$ & 2.3 & 54.4 & N & \\
16 &    33.3540840 &    36.2714462 &  5.8 & 17.152 & 16.330 & $ -13.7$ & 2.3 & 67.3 & N & \\
17 &    33.3997078 &    36.2773628 &  7.0 & 18.224 & 17.460 & $ -78.8$ & 2.3 & 45.1 & N & \\
18 &    33.4496231 &    36.2708626 &  8.3 & 19.047 & 18.278 & $-152.4$ & 2.6 & 17.8 & N & \\
19 &    33.3174171 &    36.1878624 &  0.6 & 19.387 & 18.731 & $ -59.7$ & 2.4 & 22.2 & N & \\
20 &    33.3305016 &    36.1926117 &  0.9 & 20.565 & 19.907 & $-378.7$ & 2.9 & 11.7 & Y & $1.41\pm0.21$ & $-2.3\pm0.2$\footnote{These stars are fainter that the magnitude range over which the \citet{starkenburg10} relation was calibrated; their $\FeH$ measurements should therefore be taken with cautious.} \\
21 &    33.3165016 &    36.1710815 &  0.5 & 20.667 & 20.059 & $-382.8$ & 3.1 & 10.8 & Y & $0.87\pm0.2$ & $-3.0\pm0.6^\mathrm{b}$\\
22 &    33.3029175 &    36.1470566 &  2.1 & 20.957 & 20.458 & $-387.0$ & 3.8 &  7.6 & Y & \\
23 &    33.3359184 &    36.1629181 &  1.1 & 21.109 & 20.562 & $-387.9$ & 3.6 &  7.1 & Y & \\
24 &    33.3416672 &    36.1738892 &  1.0 & 21.552 & 21.222 & $-383.1$ & 4.9 &  5.3 & Y & \\
25 &    33.3214149 &    36.1205826 &  3.5 & 21.568 & 21.196 & $-362.8$ & 5.6 &  4.5 & Y & \\
26 &    33.3535004 &    36.1727219 &  1.5 & 21.799 & 21.398 & $ -84.1$ & 8.2 &  3.5 & N & \\
27 &    33.3389587 &    36.1414452 &  2.4 & 21.814 & 21.458 & $-401.4$ & 6.6 &  4.2 & Y & \\
28 &    33.3247490 &    36.1699982 &  0.5 & 21.991 & 21.616 & $ 366.6$ & 6.9 &  3.5 & N & \\
29 &    33.3789597 &    36.1988907 &  3.0 & 22.326 & 21.877 & $-397.1$ & 7.8 &  3.1 & Y & \\
30 &    33.4619598 &    36.2307205 &  7.4 & 20.690 & 20.180 & $-215.8$ & 2.9 & 11.2 & N & \\
31 &    33.4694176 &    36.2234154 &  7.6 & 20.845 & 20.223 & $-375.8$ & 3.1 &  9.8 & Y & $3.90\pm0.45$ &$-0.7\pm0.2^\mathrm{b}$\\
32 &    33.4769173 &    36.2040291 &  7.6 & 21.132 & 20.700 & $-225.9$ & 3.5 &  6.8 & N & \\
33 &    33.2707901 &    36.0966949 &  5.5 & 17.659 & 16.706 & $ -14.2$ & 2.4 & 53.3 & N & \\
34 &    33.2869987 &    36.0956383 &  5.3 & 18.630 & 18.031 & $ -43.1$ & 2.4 & 31.0 & N & \\
35 &    33.2285843 &    36.1292763 &  5.4 & 17.306 & 16.496 & $ -44.7$ & 2.3 & 60.3 & N & \\
36 &    33.2309570 &    36.1342239 &  5.2 & 19.463 & 18.691 & $  -1.2$ & 2.4 & 21.5 & N & \\
37 &    33.2211685 &    36.1413040 &  5.4 & 17.928 & 17.216 & $ -44.2$ & 2.3 & 46.5 & N & \\
38 &    33.4219170 &    36.1992760 &  5.0 & 17.195 & 16.231 & $  10.8$ & 2.3 & 78.9 & N & \\
39 &    33.4609985 &    36.1992493 &  6.8 & 16.906 & 16.181 & $ -30.1$ & 2.4 & 57.1 & N & \\
40 &    33.3189583 &    36.1793900 &  0.2 & 17.585 & 16.692 & $-379.2$ & 2.3 & 73.3 & Y & $1.79\pm0.05$ & $-2.6\pm0.1$\\
41 &    33.3302078 &    36.1485825 &  1.8 & 17.699 & 16.858 & $ -23.2$ & 2.3 & 54.9 & N & \\
42 &    33.3511658 &    36.1708603 &  1.5 & 17.561 & 16.772 & $ -11.7$ & 2.3 & 57.2 & N & \\
43 &    33.4364586 &    36.2213593 &  6.1 & 18.337 & 17.426 & $ -93.3$ & 2.3 & 39.8 & N & \\
44 &    33.3328743 &    36.2034149 &  1.6 & 18.925 & 18.122 & $-173.5$ & 4.8 & 30.0 & N & \\
45 &    33.3890419 &    36.1674995 &  3.3 & 19.115 & 18.539 & $ -59.2$ & 2.4 & 25.2 & N & \\
46 &    33.3397484 &    36.1659431 &  1.1 & 19.286 & 18.540 & $-372.5$ & 2.4 & 26.4 & Y & $1.33\pm0.08$ & $-2.6\pm0.1$\\
47 &    33.3222084 &    36.1384430 &  2.4 & 19.019 & 18.452 & $ -71.4$ & 2.4 & 25.7 & N & \\
48 &    33.3827095 &    36.2173882 &  3.7 & 17.149 & 16.205 & $  17.3$ & 2.3 & 70.2 & N & \\
49 &    33.4824982 &    36.2344437 &  8.4 & 17.344 & 16.462 & $   2.8$ & 2.3 & 61.6 & N & \\
50 &    33.4132080 &    36.1776123 &  4.4 & 19.096 & 18.472 & $ -27.4$ & 2.4 & 26.9 & N & \\
\end{tabular}
\end{table*}

\acknowledgments

B.P.M.L. acknowledges funding through a 2012 Strasbourg IDEX (Initiative d'Excellence) grant, awarded by the University of Strasbourg. N.F.M. and B.P.M.L. gratefully acknowledge the CNRS for support through PICS project PICS06183. H.-W.R. acknowledges support by the DFG through the SFB 881 (A3). M.L.M.C. acknowledges financial support from the European Research Council (ERC-StG-335936) and from NASA through Hubble Fellowship grant \#51337 awarded by the Space Telescope Science Institute, which is operated by the Association of Universities for Research in Astronomy, Inc., for NASA, under contract NAS 5-26555. 

The data presented herein were obtained at the W.M. Keck Observatory, which is operated as a scientific partnership among the California Institute of Technology, the University of California and the National Aeronautics and Space Administration. The Observatory was made possible by the generous financial support of the W.M. Keck Foundation. The authors wish to recognize and acknowledge the very significant cultural role and reverence that the summit of Mauna Kea has always had within the indigenous Hawaiian community.  We are most fortunate to have the opportunity to conduct observations from this mountain.



\end{document}